\documentclass[preprint]{sigplanconf}

\usepackage{amsmath}
\usepackage{epsfig}

\newtheorem{example}{Example}

\newcommand{\zb}{\hspace{-0.5pt}}
\newcommand{\zbb}{\hspace{-1.0pt}}
\newcommand{\url}{}
\newcommand{\True}{\small{True}}

\begin{document}

\conferenceinfo{PEPM '10}{date, City.} 
\copyrightyear{2010} 
\copyrightdata{[to be supplied]} 

\preprintfooter{version September, 2012}   

\title{A Note on Program Specialization
}
\subtitle{ What Can Syntactical Properties of Residual Programs  Reveal? }

\authorinfo{Alexei Lisitsa}
           {{Department of Computer Science, The University of Liverpool}}
           {{A.Lisitsa@csc.liv.ac.uk}}
\authorinfo{Andrei P. Nemytykh}
           {Program Systems Institute of Russian Academy of Sciences}
           {{nemytykh@math.botik.ru}}

\maketitle

\begin{abstract}
The paper presents two examples of non-traditional using of program specialization by Turchin's supercompilation method. 
In both cases we are interested in syntactical properties of residual programs produced by supercompilation. 
In the first example we apply supercompilation to a program encoding a word equation and as a result we obtain a program  representing a graph 
describing the solution set of the word equation. 
The idea of the second example belongs to Alexandr V. Korlyukov. He considered an interpreter simulating the dynamic of the well known missionaries-cannibals puzzle. Supercompilation of the interpreter allows us to solve the puzzle. 
The interpreter may also be seen as an encoding of a non-deterministic protocol.
\end{abstract}




\section{Introduction}\label{Introduction}

One of the main concerns in the research in program specialization and other program transformation techniques is an efficiency of residual (or transformed) programs.    It is widely known also that specialization of interpreters with respect to given programs can be used for effective changing of the semantics of the programs. For example, a number of researchers were interested in generation of efficient programs implementing inverse functions 
\cite{AYuRomanenko:88}, \cite{AYuRomanenko:91}, \cite{Turchin:JFP93}, \cite{Glueck:Inverse03}. 
Still 
efficiency of the resulting programs is the major issue. 
But there is something more in program transformation techniques.  
They can be used for analysis of the programs and more specifically for the verification \cite{Leuschel:99}, \cite{Lis_Nem:IJFCS08}.

In this paper we take a step further and show that specialization can be used also not only for the analysis of the programs, but  for the solution of combinatorial and algebraic problems encoded in the programs.    We consider a number of examples illustrating use of a supercompiler (SCP4 \cite{N:03},\cite{Nemytykh:SCP4book},\cite{NT:00}) for this purpose.  
We hope the examples will be able 
to motivate further research in automated program specialization. 

\subsection{The Presentation Language}

We present our program examples in a variant of a pseudocode for functional programs 
while real supercompilation experiments with the programs were done in a strict functional programming REFAL language \cite{Turchin:Refal5}, \cite{Refal5:PZ}. 

The programs given below are written as \emph{strict} term rewriting systems based on pattern matching. The sentences in the programs are ordered from the top to the bottom to be matched. 
To be more closely to REFAL 
we use two kinds of variables: \emph{s.}variables range over \emph{symbols} (i.e. characters and identifiers, for example, \texttt{'a'} and \texttt{True}),
while \emph{e.}variables range over the whole set of the \texttt{S}-expressions. 
We also use a syntactical  sugar for representation of words (finite sequences of characters), so,  for example, the list \texttt{'b':'a':[]} is   shortened as \texttt{'ba'} and  \texttt{'aba':e.x} denotes \texttt{'a':'b':'a':e.x}.  
\texttt{<$\bullet$,$\bullet$,$\bullet$>} denotes a triple.

\section{Word Equations}\label{WordEquations}

For a given alphabet $\Sigma$ we denote as $\Sigma^*$ the set of finite words over $\Sigma$. Let $X$ be a finite set of variables disjoint with $\Sigma$. A word expression is defined to be either an element 
of $\Sigma$, a variable, the empty word $\epsilon$ or a finite sequence of word expressions. For given two word expressions $L$ and $R$, an equation of the form $L = R$ is called as a word equation. 

A solution of a word equation is any substitution of unknowns in the word equation by words that turns the equation $L = R$ 
in literal equality.

A classical problem is to describe the set of all solutions of a given word equation. In a sense the equations itself describes the set. We are interested in a more constructive (transparent) description of the set.  
Important contribution in resolving the problem was made in the late 1960's by Yu.I. Khmelevski\u{i} \cite{Khmelevskii:77}.  In 1970's G.S. Makanin \cite{Makanin:77} suggested an algorithm deciding whether there exists a solution of any given word equation or not. There exists an algorithm based on the Makanin's ideas, which for a given word equation generates a finite graph describing the corresponding solution set. 

An example given below demonstrates both the statement above 
(i.e., a solution exists)
and the fact that supercompilation is able to generate the \emph{same} graph as the Makanin's algorithm does for this example word equation. 
In this example the graph is a finite automaton. 

\noindent
\begin{verbatim}
main(e.xs) = equal('ab' ++ e.xs, e.xs ++ 'ba');

equal(s.x:e.xs, s.x:e.ys) = equal(e.xs, e.ys);
equal([], []) = True;
equal(e.xs, e.ys) = False;
\end{verbatim}

The function \texttt{equal} is a predicate checking  whether two given words are equal.  The function \texttt{main} is a predicate testing whether a given word is a solution of the following word equation:

\begin{center}
\texttt{
'ab' ++ e.xs = e.xs ++ 'ba'
}
\end{center}

\begin{figure}
\centerline{\psfig{figure=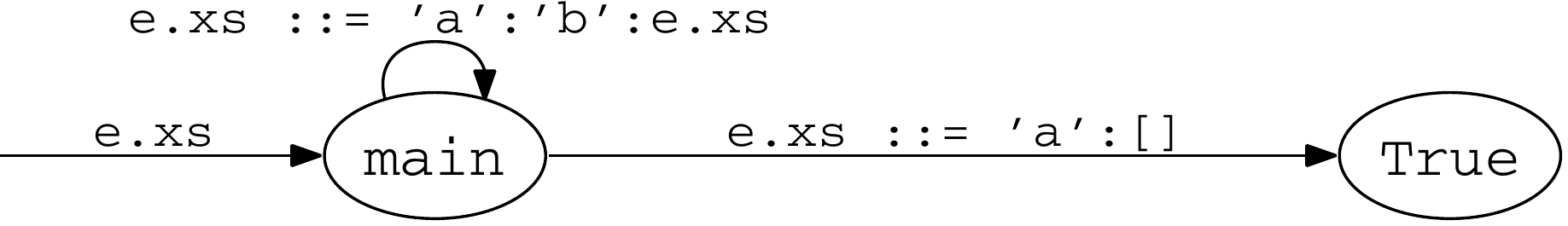,scale=0.4}}
\caption{Graph describing the solution set of the word equation.}
\label{fig:graph}
\end{figure}

We may specialize the program with respect the partial knowledge of the arguments of the call for the function \texttt{equal}. The residual program produced by the supercompiler SCP4 looks as follows:

\noindent
\begin{verbatim}
main( 'a':'b':e.xs ) = main( e.xs );
main( 'a':[] ) = True;
main( e.xs ) = False;
\end{verbatim}

This residual program \texttt{P} (as any program!) can be seen as a graph. 
Vertices of the graph correspond to the program function calls and the return expressions. Its edges are labeled with the case expressions or assignments. This graph describes the solution set of the word equation being considered. The edge corresponding to the last sentence of the \texttt{P} is redundant for such description. This sentence will be absent, if we will remove the last sentence of the original function \texttt{equal}. After that the obtained graph \emph{coincides} with the graph generated by the Makanin's algorithm (see, for example, \cite{Diekert02}). See Figure \ref{fig:graph} for the graph. The solution set is \texttt{('ab')$^*$'a'}.

Supercompilation of the following task (by the supercompiler SCP4)
\begin{verbatim}
main(e.xs) = equal('ab'++e.xs++'a',e.xs++'ba');
\end{verbatim}
yields the residual program: 

\noindent
\begin{verbatim}
main(e.xs) = False;
\end{verbatim}

The graph encoded by the residual program consists 
the root and a leaf labeled with \texttt{False}. The only edge outcoming from the root incomes in the leaf. That means the word equation
\begin{center}
\texttt{
\texttt{'ab' ++ e.xs ++ 'a' = e.xs ++ 'ba'}
}
\end{center}
has no solutions.

Resume is as follows. A result of specialization may be interesting not only because of effective performance of the residual  
program as compared  to a source program. The syntactical structure of the result may serve as a solution to the problem encoded in the source program.  
It is certainly,  that for arbitrary word equation  the supercompiler SCP4, in general,  is not able to reproduce the \emph{same} graph as the Makanin's algorithm does. Weakness of the supercompiler as well as supercompilation method \emph{per se} does not allow to achieve such a result.  
In our opinion, the ideas basing the Makanin's algorithm may be borrowed for further development of the supercompilation method.

\section{Supercompilation of an Absurd Program (a Korlyukov's Example)}\label{KorlyukovExample}

The idea of the example given in this section belongs to A.V. Korlyukov. A part of comments to the example belongs to A.V. Korlyukov as well, while another part given by the authors of this paper. 

Unfortunately, passed away prematurely A.V. Korlyukov published the idea only in Russian Internet pages \cite{Korlyukov:Missioners},\cite{Korlyukov:UserManual}.  

Following after V.F. Turchin let us consider the well-known problem ``Missionaries and Cannibals'' \cite{Turchin:Refal5}. 

\emph{Three missionaries and three cannibals come to the bank of a river and see a boat. They want to cross the river. The boat, however, can carry no more than two people. There is a further restriction. At no time should the number of cannibals on either bank of the river (including the moored boat) exceed the number of missionaries (because the missionaries would then be overpowered and eaten). How (if at all) is it possible to cross the river?}

Now we generalize the problem. Given $n$ missionaries and $k$ cannibals is it possible to cross the river and to save all the missionaries? If the answer is true, then we are interested in the algorithm carrying the strange crowd.

We intent investigate the problem in details. We will use the supercompiler SCP4 as a ``calculator'' in
an interactive mode (as usually any calculator being used). 

We consider an interpreter of the dynamic system moving the crowd from the left bank to the right. 
The interpreter is given in Figure \ref{fig:Interpreter}. 
It takes (as an input) the pair of the numbers $n, k$ describing the initial crowd on the left bank and a finite sequence of the boat states (we call such a sequence as a path). It returns information: does a prefix of the path bring the crowd to the right bank (i.e., \texttt{True} or \texttt{False})? Additionally, if the answer is \texttt{True}, then it returns the rest of the path, which did not take a part in moving the crowd; if the answer is \texttt{False}, then it returns the part of the crowd brought to the right bank.

There are a number of tricks in the encoding. We encode the pair of the numbers $n, k$ as a triple of nonnegative integers $m,p,c$ such that 
$$m = \max\{n-k,0\}, c = \max\{k-n,0\}, n = m + p, k = c + p$$ 
and we use unary notation to represent the integers. For example, five missionaries and three cannibals will be represented as \texttt{[['mm'],['ppp'],[]]}, while two missionaries and three cannibals will be represented as \texttt{[[],['pp'],['c']]}. A state of a given bank is encoded with such a triple. A state of the boat is encoded as one identifier, the name of which consists from the first capital letters of its passengers. For instance, \texttt{MM} means the state with two missionaries on the boat, while \texttt{C} is the state with one cannibal. The function \texttt{Move} has four arguments: the first one is a state of the boat, the second is an active bank (\texttt{L} or \texttt{R}), \texttt{e.l} and \texttt{e.r} are variables taking (as their values) the states of the left and right banks correspondingly. The \texttt{Move} modifies the banks' states and returns the new active bank and the two modified states. The functions \texttt{Minus} and \texttt{Plus} define unary arithmetic according to our encoding. The function \texttt{Int} iterates the crossing process. Here we skip over the functions \texttt{CutFalse, BlockRepetition} and will consider them later. As an example we give the following call for the interpreter and its result.

\noindent
\begin{verbatim}
mainInt(
 [[],['ppp'],[]],[CC,C,CC,C,MM,MC,MM,C,CC,M,MC,MC]
) 
                      = [True,[MC]]
\end{verbatim}

We would like to note that the program looks very absurd. Actually it is unthinkable to execute it at all. It is in need the whole history of its computations as the second argument to start transformations of the first argument
(though, such a program is a kind of a predicate).  
But we are not going to execute the program, we will supercompile it. And we will show that does matter. 

First of all, let us formulate the problem in the program terms. For the following parameterized call \texttt{mainInt(e.l$_0$, e.path)} and a given state \texttt{e.l$_0$} on the left bank we are interested in an answer to the question: does a path \texttt{e.path$_0$} exist such that the (a part of) result of the call is \texttt{True}?  

We will answer on the given question interactively using the supercompiler SCP4 and analyzing the residual programs produced by the supercompiler.  By the precondition of the problem we can narrow the initial left state's set 
\begin{center}
\texttt{e.left = [[e.m],[e.p],[e.c]]}
\end{center} 
to one of the following two forms  
\begin{center}
\texttt{[[e.m],[e.p],[]]} and \texttt{[[],[],[e.c]]}
\end{center}
otherwise the missionaries will immediately be eaten. Thus we have the two tasks: 
\begin{center}
\texttt{mainInt([[e.m],[e.p],[]], e.path)} 
\end{center}
and 
\begin{center}
\texttt{mainInt([[],[],[e.c]], e.path)} 
\end{center}
to be solved. 

\begin{figure}
\begin{verbatim}
mainInt(e.l, [s.a, e.path]) = 
      Int(s.a, Move(s.a,L,e.l,[[],[],[]]),e.path);

/* The boat on the left bank.  */
Move( s.a, L, e.l, e.r ) 
           = <R, Minus(s.a, e.l), Plus(s.a, e.r)>;
/* The boat on the right bank. */
Move( s.a, R, e.l, e.r ) 
           = <L, Minus(s.a, e.l), Plus(s.a, e.r)>;

Int( s.pa, R, [[],[],[]], e.r, e.path ) 
                              = True : e.path;
Int( s.pa, s.d, e.l, e.r, [] ) = False : e.r;
Int( s.pa, s.d, e.l, e.r, [] ) = CutFalse([]);
Int( s.pa, s.d, e.l, e.r, [s.pa,e.path] ) 
                       = BlockRepetition([]);
Int( s.pa, s.d, e.l, e.r, [s.x,e.path] ) = 
      Int(s.x, Move(s.x, s.l, e.l, e.r), e.path);

CutFalse( Deadlock ) = [];
BlockRepetition( Deadlock ) = [];

Minus(MM, [['mm',e.m],e.p,[]]) = [e.m,e.p,[]];
Minus(MM, [[],['pp'],[]]) = [[],[],['cc']];
Minus(MM, [['m'],['p'],[]]) = [[],[],['c']];
Minus(CC, [[],[],['cc', e.c]]) = [[],[],e.c];
Minus(CC, [e.m,['pp',e.p],[]]) 
                       = [['mm',e.m],e.p,[]];
Minus(MC, [e.m,['p',e.p],[]]) = [e.m,e.p,[]];
Minus(M, [[],['p'],[]]) = [[],[],['c']];
Minus(M, [['m',e.m],e.p,[]]) = [e.m,e.p,[]];
Minus(C, [[],[],['c',e.c]]) = [[],[],e.c];
Minus(C, [e.m,['p',e.p],[]]) = [['m',e.m],e.p,[]];

Plus(MM, [[],[],['cc']]) = [[],['mmcc'],[]];
Plus(MM, [[],[],['c']]) = [['m'],['mc'],[]];
Plus(MM, [e.m,e.p,[]]) = [['mm',e.m], e.p, []];
Plus(CC, [['mm',e.m],e.p,[]]) 
                       = [e.m, ['pp',e.p], []];
Plus(CC, [[],[],e.c]) = [[],[],['cc',e.c]];
Plus(MC, [e.m,e.p,[]]) = [e.m,['p',e.p],[]];
Plus(M, [[],[],['c']]) = [[],['p'],[]];
Plus(M, [e.m,e.p,[]]) = [['m',e.m],e.p,[]];
Plus(C, [['m',e.m],e.p,[]]) = [e.m,['p',e.p],[]];
Plus(C, [[],[],e.c]) = [[],[],['c',e.c]];
\end{verbatim}
\caption{The interpreter.}
\label{fig:Interpreter}
\end{figure}

With the goal to make the residual programs more compact (so easier analyzable) we add an additional trick. Among all the paths, moving the boat from one bank to the other, there exist meaningless. 
The simplest from them contains at least two identical states staying side by side in the path: such two states mean the boat has just moved passengers to a bank immediately moves the same kind of passengers back to the other bank. If there exists such a path successfully moving the crowd to the right bank, then there exits a shorter successful path, where the conjugated states are removed. 
This note allows us 
to 
exclude the indicated kind of the paths (henceforth everywhere in this paper).  We do that by calling the function \texttt{BlockRepetition} with an argument which never matches the function definition. So the call leads to an abnormal stop at run time and to cutting the corresponding branch of the program at supercompile time. The first argument \texttt{s.pa} of the function \texttt{Int} serves to recognize two identical states staying side by side in a given path.

Whenever we hope on the positive answer to the problem question we will replace the following sentence of the program to be supercompiled 

\noindent
\begin{verbatim}
Int( s.pa, s.d, e.l, e.r, [] ) = False : e.r;
\end{verbatim}
\noindent
with the sentence

\noindent
\begin{verbatim}
Int( s.pa, s.d, e.l, e.r, [] ) = CutFalse([]);
\end{verbatim}

Once again we do that for the sake of simplicity of the corresponding residual programs. The call \texttt{CutFalse([])} like the call for the function \texttt{BlockRepetition} cuts the branch leading to the negative answer.

\subsection{Diagonal Cases}

We start with the case when the number of the missionaries equals to the number of the cannibals. That is to say, we have to supercompile the following start configuration: 
\begin{center}
\texttt{mainInt([[],[e.p],[]], e.path)}
\end{center}

Our first experiment is for two missionaries and two cannibals.

\begin{example}\label{Example1}

The start configuration: 
\begin{center}
\texttt{mainInt([[],['pp'],[]], e.path)}
\end{center}
The program contains the sentence with the call \texttt{CutFalse([])}.
Supercompilation produces the following residual program:

\noindent
\begin{verbatim}
mainInt'([s.x , e.path]) = True:[f(s.x,e.path)];

f(CC, [C,MM,M,MC,e.path]) = e.path;
f(CC, [C,MM,C,CC,e.path]) = e.path;
f(CC, [C,M,MC,s.x,e.path]) = f(s.x, e.path);
f(MC, [M,MM,M,MC,e.path]) = e.path;
f(MC, [M,MM,C,CC,e.path]) = e.path;
f(MC, [M,C,CC,s.x,e.path]) = f(s.x, e.path);
\end{verbatim}

The exits from the recursion show the length of the shortest path moving the crowd to the right bank is 5. There exist four such paths:

\noindent
\begin{verbatim}
[CC, C, MM, M, MC]
[CC, C, MM, C, CC]
[MC, M, MM, M, MC]
[MC, M, MM, C, CC]
\end{verbatim}

Moreover, the residual program specifies the whole set of the successful paths.

\end{example}

We leave to the reader to analyze the residual programs for one missionary and one cannibal (Example \ref{Example2}) and for three missionaries and three cannibals (Example \ref{Example3}). The source programs contain the sentence with the call \texttt{CutFalse([])}. 
The answers to the question are positive in both cases.

\begin{example}\label{Example2}
\noindent
\begin{verbatim}
mainInt'([MC , e.path]) = True : [e.path];
\end{verbatim}
\end{example}

\begin{example}\label{Example3}
\noindent
\begin{verbatim}
mainInt'([s.x , e.path]) = True : [f(s.x,e.path)];

f(CC,[C,CC,C,MM,MC,MM,C,CC,M,MC,e.path]) = e.path;
f(CC,[C,CC,C,MM,MC,MM,C,CC,C,CC,e.path]) = e.path;
f(CC,[C,M,MC,s.x,e.path]) = f(s.x, e.path);
f(MC,[M,CC,C,MM,MC,MM,C,CC,M,MC,e.path]) = e.path;
f(MC,[M,CC,C,MM,MC,MM,C,CC,C,CC,e.path]) = e.path;
f(MC,[M,C,CC,s.x,e.path]) = f(s.x, e.path);
\end{verbatim}

\end{example}

The following experiment is for an equal number of missionaries and cannibals, and the number 
is greater than three.

\begin{example}\label{Example4}

The start configuration: 
\begin{center}
\texttt{mainInt([[],['pppp', e.p],[]], e.path)}
\end{center}
The residual program for the source program containing the sentence with \texttt{False} is given in the attachment \ref{Appendix}. The residual program never returns \texttt{True} and that is a \emph{syntactical} property of the program. On the other hand,  the source program terminates for \emph{any given} arguments. That allows us to conclude that there exist no paths moving the crowd to the right bank. 

Note that if we supercompile the source program containing the sentence with the call \texttt{CutFalse([])}, then the residual program is \emph{empty}. In other words, it is a trivial program with the empty domain and this property also is syntactical rather than semantic. We might infer the negative answer to the problem question from the emptiness of the residual program.

\end{example}

Thus we have considered all the diagonal cases. If the number of both missionaries and cannibals is greater than three, then there exist no paths moving the crowd to the right bank, otherwise such a path exists.

\subsection{The Number of Missionaries is Greater Than the Number of Cannibals}

Now we consider the case when the number of the missionaries greater 
than the number of the cannibals. The corresponding start configuration is: 
\begin{center}
\texttt{mainInt([['m',e.m],[e.p],[]], e.path)}
\end{center}

In all our experiments concerning to this case the source programs contain the sentence with the call \texttt{CutFalse([])}.

\begin{example}\label{Example5}

The start configuration is 
\begin{center}
\texttt{mainInt([['m',e.m],[e.p],[]], e.path)}
\end{center}
The supercompiler SCP4 produces 
quite a large residual program. The program may be found in \cite{Lis_Nem:MissionariesAttachment}. Unfortunately we have to analyze the residual program, because the numbers of both missionaries and cannibals are not given in advance, and for each such a pair may exists a needed path and such paths may have (and do have) 
very different structures. Moreover, the method used in the first example cannot be applied here, because it is unknown in advance to which pair of the numbers a given exit from the recursion corresponds, and maybe for a pair there exist no exits from the recursion at all (i.e., all paths starting with a given pair lead to an abnormal stop (a deadlock) of the program). 

\end{example}

On the other hand, such a large residual program may be an indirect evidence of the fact that the set of the successful paths is too large. 
Assiming that we may try to narrow the paths' set, where we are looking for the successful path's witnesses. With such an aim we restrict the boat states as follows. 

\begin{example}\label{Example6}

The boat crossing the river in the direction to the right bank may have only three states \texttt{MM, MC, MC}, while it crossing the river in the opposite direction may have only two states \texttt{M, C}. The idea is to increase stepwise the number of the people on the right bank and in such a way to approach to resolution of the task. We do that by changing the original function \texttt{Move} with the following:

\noindent
\begin{verbatim}
/* The boat on the left bank. */
Move( MM, L, e.l, e.r ) 
             = <R, Minus(MM, e.l), Plus(MM, e.r)>;
Move( MC, L, e.l, e.r ) 
             = <R, Minus(MC, e.l), Plus(MC, e.r)>;
Move( CC, L, e.l, e.r ) 
             = <R, Minus(CC, e.l), Plus(CC, e.r)>;

/* The boat on the right bank. */
Move( M, R, e.l, e.r ) 
             = <L, Minus(M, e.l), Plus(M, e.r)>;
Move( C, R, e.l, e.r ) 
             = <L, Minus(C, e.l), Plus(C, e.r)>;
\end{verbatim}

This function filters out the forbidden states of the boat: they never match with the sentences of the redefined function \texttt{Move} and they lead to an abnormal stop (recognition impossible, deadlock).

Now, supercompiling the start configuration 
\begin{center}
\texttt{mainInt([['m'],[ 'pp'],[]], e.path)}
\end{center}
we obtain the following residual program.

\noindent
\begin{verbatim}
mainInt'([MC, M, MM, M, MM, M, MC, e.path]) 
                              = True : [e.path];
mainInt'([MC, M, MM, M, MM, C, CC, e.path]) 
                              = True : [e.path];
mainInt'([MC, M, MM, M, MC, C, MC, e.path]) 
                              = True : [e.path];
mainInt'([MC, C, MC, M, MM, M, MC, e.path]) 
                              = True : [e.path];
mainInt'([MC, C, MC, M, MM, C, CC, e.path]) 
                              = True : [e.path];
mainInt'([MC, C, MC, M, MC, C, MC, e.path]) 
                              = True : [e.path];
mainInt'([CC, C, MM, M, MM, M, MC, e.path]) 
                              = True : [e.path];
mainInt'([CC, C, MM, M, MM, C, CC, e.path]) 
                              = True : [e.path];
mainInt'([CC, C, MM, M, MC, C, MC, e.path]) 
                              = True : [e.path];
\end{verbatim}

Analyzing this result program we see that if the boat is on the left bank, then the path \texttt{[MC,C,MC,M]}  adds a pair (missionary-cannibal) on the right bank, provided that on the left bank the number of the missionaries greater than the number of the cannibals. That means a repeated iteration of such a path leads to success. The path \texttt{[MC,C,MC]} (a prefix of the considered above) decides the problem with two missionaries and one cannibal.

Note that such a narrowing of the possible paths might be unsuccessful for other start configurations. For example, in the case of the start configuration
\begin{center}
\texttt{mainInt([[],['ppp'],[]], e.path)}
\end{center}
 there exists no 
 successful path from the narrowed set, while we have found a successful path in the whole paths' set (see Example \ref{Example3}). 
 
 Nevertheless supercompilation of the start configuration 
\begin{center}
\texttt{mainInt([['m'],['pp'],[]], e.path)} 
\end{center}
with the original definition of the function \texttt{Move} produces a residual program \texttt{P} (see the attachment \cite{Lis_Nem:MissionariesAttachment}), which is more complicated 
than 
the residual program above, but the key information of the path \texttt{[MC,C,MC,M]} still may be retrieved from the \texttt{P} (but applying more efforts).

\end{example}

The other cases (when in the start configurations the number of the cannibals is greater than the number of the missionaries, no cannibals, no missionaries) are trivial.
 The following table summarizes our investigation.  It gives a left upper angle of the result matrix. The empty cells correspond to the \texttt{False} answer.

\begin{center}
  \begin{tabular}{l|c|c|c|c|c|c|c|c}
    \hfill \zbb\zbb\zb\small{M.\tiny{$\rightarrow$}}  & 0  & 1  & 2  & 3 & 4 & 5 & 6 & 7  \\
    \zbb\zbb\zb\small{C.\tiny{$\downarrow$}}          &    &    &    &   &   &   &   &    \\
    \hline
    0                     & \True & \True & \True & \True & \True & \True & \True & \True \\
    \hline
    1                     & \True & \True & \True & \True & \True & \True & \True & \True \\
    \hline
    2                     & \True &       & \True & \True & \True & \True & \True & \True \\
    \hline
    3                     & \True &       &      & \True & \True & \True & \True & \True \\
    \hline
    4                     & \True &       &      &       &       & \True & \True & \True \\
    \hline
    5                     & \True &       &      &       &       &       & \True & \True \\
    \hline
    6                     & \True &       &      &       &       &       &       & \True \\
    \hline
    7                     & \True &       &      &       &       &       &       &  \\
  \end{tabular}
\end{center}

\section{Conclusion}

What did happen in the experiments described above? Program specialization (supercompilation) originally developed as an optimization method was used for studying some \emph{syntactical} properties of the residual programs rather than for executing the programs. In a sense we used the supercompiler SCP4 as a tool helping to solve the mathematical tasks. We used the tool in an interactive fashion. In such a way we use any calculator.

In the section \ref{WordEquations} we were interested in the graph encoded 
within the residual program. The graph describes the solution set of a word equation. 

Using the supercompiler SCP4 to solve the missionaries-can\-ni\-bals problem was more complicated. In this case, using a rough analogy, the supercompiler was exploited as a kind of a ``PROLOG interpreter''. We were interested in narrowing a parameter of the start configurations (the \emph{goals} to be supercompiled) (\texttt{mainInt([[e.m],[e.p],[e.c]], e.path)}). The narrowed parameter \texttt{e.path} (its \emph{substitution set}) has to satisfy a property imposed by the source program (actually a kind of a predicate). Unlike PROLOG, when the substitution set cannot be described as a finite union of parameterized lists (S-expressions), the supercompiler produces a finite residual program describing the substitution set more explicitly (transparently) as to compared with the source program. (The same note is valid for the word equation example.)  The function calls \texttt{CutFalse([]), BlockRepetition([])}are analogues of the CUT mechanism in PROLOG. 

The function \texttt{mainInt(e.l,e.path)} can be seen as an interpreter transforming data \texttt{e.l$_0$} according to a given program \texttt{e.path$_0$}. In such a case our experiments can be seen as specialization of the interpreter with respect to partial information of the data. The authors used such specialization of interpreters for verification of \emph{safety} properties of nondeterministic cache coherence protocols \cite{Lis_Nem:Programming}, \cite{Lis_Nem:IJFCS08}, \cite{Lis_Nem:CSR07}, \cite{Lis_Nem:Protocols07}. The dynamic system ``missionaries-cannibals'' also may be considered as a nondeterministic protocol, where the boat states are actions controlling the global state of the computing system. The actions are being chosen in a nondeterministic way governed by a number of guides. The examples above of the start crowd configurations, for which we proved the negative answer on the question given in the problem, are the conditions on the start protocol configurations satisfying the safety property defined by the positive answer on the question. That is to say, the computer system starting with such a configuration never reaches the global state, when the whole crowd is on the right bank. Our treatment of the start configurations corresponding to the positive answer is a search for a witness violating the safety property. Acting in a similar way we have found bugs in a number of cache coherence protocols (see \cite{Lis_Nem:TAP08}, \cite{Lis_Nem:Protocols07}).

{Finally we would like to note that actually all our experiments were done in a strict functional programming language REFAL \cite{Turchin:Refal5},\cite{Refal5:PZ}, using the supercompiler SCP4 under the following strategy:
\noindent
\begin{verbatim}
$MATCHING ForRepeatedSpecialization;
$STRATEGY Applicative;
\end{verbatim}
}

\appendix
\section{The Residual Program Obtained in Example \ref{Example4}}\label{Appendix}

\noindent
\begin{verbatim}
mainInt'(e.p, [s.x : e.path]) 
               = False : [[],f(s.x,e.path,e.p)];

f(CC, [], e.p) = <[], ['cc']> ;
f(CC, [C], e.p ) = <[],['c']> ;
f(CC, [C, CC], e.p) = <[],['ccc']> ;
f(CC, [C, CC, C], e.p) = <[],['cc']> ;
f(CC, [C, CC, C, MM], e.p) = <['pp'],[]> ;
f(CC, [C, CC, C, MM ,MC], e.p) = <['p'],[]> ;
f(CC, [C, CC, C, CC, e.path], e.p) 
              = <[], ['ccc', g(e.p, [], e.path)]>;
f(CC, [C, M], e.p) = <['p'],[]> ;
f(CC, [C, M, MC], e.p) = <[],[]> ;
f(CC, [C, M, MC, s.x, e.path], e.p) 
                       = f(s.x,e.path,e.p);
f(MC, [], e.p) = <['p'],[]> ;
f(MC, [M], e.p) = <[], ['c']> ;
f(MC, [M, CC], e.p) = <[], ['ccc']> ;
f(MC, [M, CC, C], e.p) = <[], ['cc']> ;
f(MC, [M, CC, C, MM], e.p) = <['pp'], []> ;
f(MC, [M, CC, C, MM, MC], e.p) = <['p'], []> ;
f(MC, [M, CC, C, CC, e.path], e.p) 
              = <[], ['ccc', g(e.p, [], e.path)]>;
f(MC, [M, C], e.p) = <[], ['cc']> ;
f(MC, [M, C, CC], e.p) = <[], []> ;
f(MC, [M, C, CC, s.x, e.path], e.p)
                       = f(s.x,e.path,e.p);
f(C, [], e.p) = <[], ['c']> ;


g(e.p, e.c, []) = 'c':e.c;
g(e.p, e.c, C) = e.c;
g('p':e.p, e.c, [C, CC, e.path]) 
               = g(e.p, 'c':e.c, e.path); 

\end{verbatim}

\bibliographystyle{abbrvnat}

\end{document}